\documentstyle[11pt,aaspp4,psfig]{article}
\def\k{\mbox{\boldmath$k$}}
\def\q{\mbox{\boldmath$q$}}
\def\x{\mbox{\boldmath$x$}}
\def\v{\mbox{\boldmath$v$}}
\def\ALP{\mbox{\boldmath$\alpha$}}
\def\BET{\mbox{\boldmath$\beta$}}
\def\GAM{\mbox{\boldmath$\gamma$}}
\def\EPS{\mbox{\boldmath$\epsilon$}}

\def\A{\mbox{\boldmath$A$}}
\def\B{\mbox{\boldmath$B$}}
\def\C{\mbox{\boldmath$C$}}
\def\D{\mbox{\boldmath$D$}}
\def\E{\mbox{\boldmath$E$}}

\def\dm{\delta_m}
\def\dg{\delta_g}
\def\Dm{\Delta_m}
\def\Dg{\Delta_g}

\slugcomment{\hfill RESCEU-30/99, UTAP-342} 
\begin{document}
%
%
\title{\LARGE\bf Stochastic Biasing and Weakly Non-linear Evolution of 
Power Spectrum} 
%
%
%
%
\author{Atsushi Taruya} 
\affil{
  Research Center for the Early Universe (RESCEU), 
  School of Science, University of Tokyo, 
  Tokyo 113-0033, Japan 
}
\authoraddr{E-mail: ataruya@utaphp1.phys.s.u-tokyo.ac.jp}
\begin{abstract} 
Distribution of galaxies may be a biased tracer 
of the dark matter distribution and the relation between the 
galaxies and the total mass may be stochastic, non-linear and 
 time-dependent. Since many observations of galaxy 
 clustering will be done at high redshift, 
 the time evolution of non-linear stochastic biasing 
 would play a crucial role for the data analysis of the future 
 sky surveys. Recently, analytic study of the time evolution 
 induced by gravity has been reported in the mildly non-linear regime. 
 Here, we further develop the non-linear analysis including the 
 next-to-leading order and attempt to clarify 
 the non-linear feature of the stochastic biasing. 
 Employing the perturbative approach, 
 we compute the one-loop correction of the power spectrum for the   
total mass, galaxies and their cross correlation. 
Assuming that the initial distribution of galaxies is given by the 
local function, we specifically investigate   
the time evolution of the biasing parameter 
and the correlation coefficient deduced from the power spectra. 
On large scales, we find that the time evolution of the biasing parameter 
could deviate from the linear theory prediction in 
 presence of the initial skewness, even though the scale-dependence 
of the biasing is very weak. On the other hand, 
 the deviation can be reduced if the stochasticity between 
 the galaxies and the total mass exists. 
To explore the influence of non-linear gravity, we focus on the 
 quasi-linear scales, 
 where the non-linear growth of the total mass becomes important. 
 It is recognized that the scale-dependence of the biasing 
dynamically appears and the initial stochasticity could affect 
the time evolution of the scale-dependence. 
The result is compared with the recent N-body simulation that 
the scale-dependence of the halo biasing can appear on relatively 
large scales and the biasing parameter takes the lower value on smaller scales. 
Qualitatively, our weakly non-linear results can explain this trend 
if the halo-mass biasing relation has the large scatter at high redshift. 
\end{abstract}
%
%
%
\section{Introduction}
\label{sec: intro}
%
%
%
%
%
%
%
Understanding the large scale structure of our universe is 
       the ultimate goal of observational cosmology. Currently,  
       Sloan Digital Sky Survey and Two-Degree Field 
       Survey are working and the enormous observational data 
       will be obtained. In the next generation of deep galaxy 
       surveys, the formation and the evolution of galaxy distribution will 
       be measured more precisely. It provides us with the powerful constraint 
       on the theory of structure formation in the Universe.  
        
       When we compare the theoretical prediction 
       for the density fluctuations with the observation of galaxy sky 
       survey, however, the relation between the total mass and the 
       galaxies must be clarified. Since we do not yet have any reliable 
       theory of galaxy/star formation history, there exists  
       the statistical uncertainty between the galaxies and the 
       total mass, which must be determined from the 
        observational data. This uncertainty hampers the effort to 
        obtain the cosmological 
       information such as the density parameter (\cite{hamilton}). 

       Provided the fluctuations of the total mass $\dm$ and the 
       distribution of the galaxy number density $\dg$, the relation 
\begin{equation}
 \dg=b~\dm
\label{linear-bias}
\end{equation}
       is often quoted in the literature. However, 
       the galaxy biasing in nature could be non-linear and the 
       simple linear relation (\ref{linear-bias}) might not be 
       appropriate (\cite{FG93}). Fry (1994) and Gazta\~naga \& 
       Frieman (1994) compared the observational data with  
       a straightforward extension of (\ref{linear-bias}) given by 
       \begin{equation}
 \dg=f(\dm)=\sum_{n=1}\frac{b_n}{n!}~(\dm)^n.
        \label{nl-bias}
       \end{equation}
       The assumption (\ref{nl-bias}) still restricts the statistical  
       feature of the galaxy distribution. In general, 
       the relation between the galaxies and the total mass may be 
       time dependent and stochastic (\cite{CO92}). Therefore, 
       a more general framework to treat this situation is 
       needed, referred to as the stochastic biasing. 

       Using the different catalogs or subsamples of 
       the galaxy surveys, 
       the observational evidence of stochasticity 
       has been recently discussed (\cite{TB99}, \cite{Seaborne99}). 
As for the theoretical study, Dekel \& Lahav (1999) has proposed 
       the general formalism for non-linear stochastic biasing.    
       Pen (1998) discussed an observational method to 
       determine the stochasticity. Further, 
       Blanton {\it et al}. (1998) and Blanton, Cen \& Tegmark (1999)  
       performed the cosmological simulation of galaxy formation and  
       demonstrated that the stochastic property of the biasing could 
       evolve due to the galaxy formation process. 
       Since we still lack our knowledge of the 
       non-linear aspect of the stochastic biasing,   
       the qualitative features should deserve consideration .

In this paper, we analytically investigate the 
        dynamical feature of the non-linear stochastic biasing.   
        We study the time evolution induced by gravity, 
      which would be one of the most important sources of 
        non-linearity. 
      The observational study of the galaxy clustering 
      recently shows that the evolution model induced by gravity is 
      sufficient to explain the observed galaxy distribution 
      at redshift $z<2$ (\cite{MMLW99}, \cite{MBML99}). 
      Therefore, the gravitational evolution of biasing would play a 
      crucial role in the data analyses of future deep sky surveys. 

       The previous discussion about the time evolution of 
       stochastic biasing has been concentrated on the tree-level 
       analysis, i.e, based on the lowest order perturbation.  
       Tegmark \& Peebles (1999) examined the linear evolution. 
       Taruya, Koyama \& Soda (1999) has extended 
       the pioneering work of Fry (1996) and 
       investigated the skewness and the bi-spectrum. 
       Taruya \& Soda (1999) further study the 
       weakly non-linear galaxy-mass density relation. 

       This paper focuses on the next-to-leading order of non-linearity. 
       We take into account the one-loop correction of non-linear gravity,  
       corresponding to the one-loop diagram in the graphical 
       representation of perturbation theory (\cite{SF96a}, 1996b).     
       Then, we shall investigate the power spectrum, 
       the second order statistics of density perturbations 
       in the Fourier space. 
       The power spectrum of the total mass distribution  
       has been extensively investigated by several authors 
       (\cite{MSS92}, \cite{JB94}, \cite{BE94}, \cite{SF96b}). 
        Our analysis is further extended to the power spectrum of 
        galaxies and the cross correlation between $\dm$ and $\dg$.  
        Define the vector 
        $\hat{X}^{\dagger}=(\hat{\delta}_m(\k),\hat{\delta}_g(\k))$ 
        in the Fourier space, we have the three kinds of power spectrum 
        \begin{equation}
 \langle\mbox{$\hat{X}$}(\k_1)\mbox{$\hat{X}$}^{\dagger}(\k_2)\rangle
  =(2\pi)^3\delta_D(\k_1+\k_2)~ \left(
\begin{array}{cc}
P_m(k_1) & P_{\times}(k_1)\\
P_{\times}(k_1) & P_g(k_1)\\
\end{array}
\right),
\label{spectra}
       \end{equation}
       where $P_m(k)$, $P_g(k)$ and $P_{\times}(k)$  are the 
       power spectrum of the total mass, galaxies and the cross 
       correlation between them. The statistical feature of 
       the galaxy biasing is encoded 
       in the power spectra $P_g(k)$ and $P_{\times}(k)$, which can be 
       quantified by comparing them with the power spectrum of the total 
       mass distribution. We will analyze the biasing parameter $b_k$ and the 
       correlation coefficient $r_k$ : 
\begin{equation}
 b_k \equiv \sqrt{\frac{P_g(k)}{P_m(k)}},~~~~~~~~~~~
 r_k \equiv \frac{P_{\times}(k)}{\sqrt{P_g(k)P_m(k)}},     
\label{def-of-b-r2}  
\end{equation}
 which are frequently used in the literature (\cite{TP98},
 \cite{TB99}). Note that the simple biasing prescription 
 (\ref{linear-bias}) restricts the above parameter $r_k=1$. 
 
We organize the paper as follows;   
In section \ref{sec: Formalism},  we briefly mention 
the basic formalism for the time evolution of stochastic biasing. 
The evolution equations (eqs.[\ref{basic-eq1}][\ref{basic-eq2}]
[\ref{basic-eq-gal}]) and the initial conditions 
(eqs.[\ref{init-condition}][\ref{init-g}]) are presented.
Section \ref{sec: evolution} is devoted to 
the weakly non-linear analysis. We compute the one-loop correction of 
the power spectra and the time evolution of the stochastic biasing is 
investigated in detail. Our main results are  
described in Section \ref{subsec: time evolution}. 
We specifically pay attention to  
the effect of the non-linear biasing (Sec.\ref{subsubsec: linear}) 
and the non-linear gravity (Sec.\ref{subsubsec: quasi-linear}). 
The conclusions and discussion are 
summarized in section \ref{sec: conclusion}. 
%
%
%
%
\section{Gravitational evolution of Stochastic biasing} 
\label{sec: Formalism}
%
%
%
Recall the definition of the density field $\dm$ and $\dg$: 
\begin{equation}
 \dm(x)\equiv\frac{\rho(x)-\bar{\rho}}{\bar{\rho}},~~~~~~~~~
 \dg(x)\equiv\frac{n_g(x)-\bar{n}_g}{\bar{n}_g}, 
\end{equation}
where the variables with overbars denote the homogeneous averaged
quantity. 

%
The gravitational evolution of the 
  galaxy biasing has been studied 
   by Fry (1996), Tegmark \& Peebles (1998) and 
   Taruya, Koyama \& Soda (1999). 
   In these papers, the distribution $\dm$ follows 
  the equation of continuity and the Euler equation. On the 
  other hand, neglecting the merging and the galaxy formation process, 
  the galaxies are regarded as the test particles in the 
  dark matter distribution $\dm$. Thus, the distribution $\dg$ 
  should satisfy the equation of continuity whose velocity field is 
  determined by the gravitational potential. 
  On large scales, the assumption of the irrotational velocity 
  flow is valid. We define the velocity divergence 
  $\theta\equiv\nabla\cdot\v/(aH)$, where $a$ is the scale 
  factor of the universe and $H$ is the Hubble parameter.  
  The evolution equations for the total mass and the galaxies 
  become 
\begin{eqnarray}
&& \frac{\partial\delta_m}{\partial t}+H\theta+
\frac{1}{a}\nabla\cdot(\delta_m\v)=0,
\label{basic-eq1}
\\
&&  \frac{\partial\theta}{\partial t}+\left(1-\frac{\Omega}{2}+
\frac{\Lambda}{3H^2}\right)H\theta+\frac{3}{2}H\Omega\delta_m
+\frac{1}{a^2H}\nabla\cdot(\v\cdot\nabla)\v=0,
\label{basic-eq2}
\end{eqnarray}
\begin{equation}
  \frac{\partial\delta_g}{\partial t}+H\theta+
\frac{1}{a}\nabla\cdot(\delta_g\v)=0.
\label{basic-eq-gal}
\end{equation}
The variable $\Lambda$ is the cosmological constant and $\Omega$ is the 
density parameter defined by 
\begin{equation}
\Omega\equiv \frac{8\pi G}{3}\frac{\bar{\rho}}{H^2}.  
\end{equation}
%

Equations (\ref{basic-eq1}), (\ref{basic-eq2}) and  
  (\ref{basic-eq-gal}) are 
  our basic equations for the time evolution of $\dm$ and $\dg$. 
  These equations can be solved systematically using the 
  perturbation theory if we        
  give the initial conditions $\delta_{m,g}(x,t_i)$ at an early 
  time $t_i$ order by order. 
  To linear order, provided 
  the initial distributions $\dm$ and $\dg$ are given in terms of 
  the independent random fields $\Dm(x)$ and $\Dg(x)$, the solutions 
  become 
\begin{equation}
 \hat{\delta}_m^{(1)}(x)=D(t)~\Delta_m(x),~~~~~~~~~~
  \hat{\delta}_g^{(1)}(x)=\left\{D(t)-1\right\}\Delta_m(x)
+\Delta_g(x). 
\label{linear-sol}
\end{equation}
Since we are interested in the leading behavior of the time evolution, 
the solutions are only taken into account the growing mode of perturbation 
denoted as $D(t)$. The function $D(t)$ satisfies   
\begin{equation}
 \ddot{D}+2H \dot{D}-\frac{3}{2}H^2\Omega D=0, 
  \label{linear-perturbation}
\end{equation} 
with the initial condition $D(t_i)=1$. Note that the random 
fields $\Dm$ and $\Dg$ assign the statistical feature of the 
galaxies and the total mass. On large scales, 
the linear perturbation for the fluctuations $\delta_{m,g}^{(1)}$ 
may be valid and the central limit theorem implies that the spatial 
distribution could be well-approximated by the 
Gaussian statistics. Hence, the variables $\Dm$ 
and $\Dg$ are assumed to be the Gaussian random fields. 

The full evolution is non-linear, including the higher order 
      perturbative solutions. Regarding the spatial variables 
      $\Delta_{m,g}$ as the seeds of perturbation, 
      the general form of the $n$-th order solutions 
      becomes the terms collecting the time dependent function 
      multiplied by the spatial random variables $[\Delta_{m,g}]^n$. 
      Here, we give the initial distributions $\delta_{m,g}$ 
      as the Taylor expansion of the functions (\cite{TKS99}, \cite{TS99})
\begin{equation}
\dm(x,t_i)=f(\Dm),~~~~~~~~~\dg(x,t_i)=g(\Dg).   
 \label{init-condition}
\end{equation}
     Based on the ansatz (\ref{init-condition}), 
     we will consider the time evolution emanating from the local biasing. 
     Their functional forms are determined as follows; 
since the total mass fluctuation is produced during 
      the very early stage of the universe, the gravitational 
      instability induces 
      the deviation from the initial Gaussian distribution. 
      Thus, we give the initial condition of the total mass $f(\Dm)$ 
      iteratively from the perturbative solution by dropping the 
      decaying mode. As for the galaxies, the initial 
      distribution $\dg$ is induced by the galaxy formation. 
      Currently, it is not feasible to know 
      the initial non-Gaussian distribution 
      $g(\Dg)$. We treat $g$ as a parametrized function, 
\begin{equation}
 g(\Dg)=\Dg+\frac{h_1}{6}(\Dg^2-\langle\Dg^2\rangle)+
  \frac{h_2}{24}\Dg^3+\cdots.
\label{init-g}
\end{equation}
      Neglecting the higher order terms,  (\ref{init-g}) 
      gives the same initial condition as in the linear order 
      solution $\dg^{(1)}$. 

      Notice that the resulting expression for the total mass $\dm$ 
      leads to the non-local and non-Gaussian distribution 
      (see eqs.[\ref{2nd-sol-m}][\ref{3rd-sol-m}]). On the other hand, 
      we simply assume that the function $g(\Dg)$ is locally 
      characterized by the constant parameters $h_1$ and $h_2$,  
      which have to do with the initial skewness and initial kurtosis. 
      We should remark that the locality of the galaxy distribution 
      only holds at the initial time. 
      Due to the gravity, the distribution $\dg$ 
      can become non-local, which can yield  
      the scale-dependence of the galaxy biasing (Sec.
      \ref{subsubsec: quasi-linear}, see also \cite{M99}). 

      Once we obtain the perturbative solutions $\delta_{m,g}$, 
      we can investigate the statistical features for the density 
      fluctuations by taking the ensemble average
      $\langle\cdots\rangle$. In our prescription, the ensemble average 
      is taken with respect to the Gaussian random fields $\Delta_{m,g}$. 
      Their stochasticity is completely characterized by the 
      three quantities. Defining the two-dimensional vector 
      $\mbox{$\hat{Y}$}^{\dagger}=(\hat{\Delta}_m, \hat{\Delta}_g)$ 
      in the Fourier space, we have 
\begin{eqnarray}
 \langle\mbox{$\hat{Y}$}(\k_1)\mbox{$\hat{Y}$}^{\dagger}(\k_2)\rangle
  =(2\pi)^3\delta_D(\k_1+\k_2)~P_l(k_1) \left(
\begin{array}{cc}
1 & b_0~r_0\\
b_0~r_0 & b_0^2\\
\end{array}
\right),
\label{initial-spectra}
\end{eqnarray}
       where the quantity $P_l(k)$ is the initial power spectrum 
       of the total mass. The parameters $b_0$ and $r_0$ are assumed to 
       be the scale-independent variables, for simplicity.   
%

Now, the formalism developed here enables us to obtain the 
weakly non-linear power spectra (\ref{spectra}). 
Before closing this section, we present the 
      linear order results. 
       Using the relation (\ref{initial-spectra}), 
       the three kinds of the power spectra are obtained by 
       substituting the Fourier transform of the linear solutions 
       (\ref{linear-sol}) into the definition (\ref{spectra}). 
       Denoting the linear spectrum as $P^{(11)}(k)$, 
       we have 
\begin{eqnarray}
 P_m^{(11)}(k)&=&D^2~P_l(k),
\nonumber\\
 P_g^{(11)}(k)&=&\left\{(D-1)^2+2 b_0 r_0 (D-1)+b_0^2\right\}~P_l(k),
\label{linear-PS}\\
 P_{\times}^{(11)}(k)&=& D\left\{(D-1)+b_0 r_0\right\}~P_l(k).
\nonumber
\end{eqnarray}
Thus, the biasing parameter $b_k$ and the correlation coefficient $r_k$ 
can be evaluated from (\ref{def-of-b-r2}) (\cite{TP98}, \cite{TKS99}): 
\begin{equation}
 b^{(l)}(t)=\frac{\sqrt{(D-1)^2+2 b_0 r_0 (D-1) + b_0^2}}{D},~~~
 r^{(l)}(t)=\frac{1}{b^{(l)}}~\frac{D-1+b_0 r_0}{D}, 
\label{linear-b-r}
\end{equation} 
where the superscript $(^{(l)})$ means the linear order quantity. 
The variables $b^{(l)}$ and $r^{(l)}$ are 
scale-independent and their initial values correspond to the parameters 
$b_0$ and $r_0$. Note that Fry (1996) has obtained 
the same result in the specific case $r_0=1$.  
In this case, we immediately obtain $r^{(l)}=1$ and the relation between 
$\dm$ and $\dg$ reduces to the deterministic biasing 
(\ref{linear-bias}). 
       Thus, in our prescription of the time evolution, 
       the stochasticity in the galaxy biasing comes from the 
       initial condition $r_0<1$.          
%
%
%
%
%
%
%
%
%
%
%
%
%
\section{Perturbative approach to the non-linear power spectra} 
\label{sec: evolution}
%
%
%
%
In this section, we develop the higher order perturbations and 
investigate the time evolution of galaxy biasing. 
In section \ref{subsec: 1-loop}, 
the one-loop correction of power spectra is computed. 
After describing the qualitative feature of the galaxy biasing 
in the weakly non-linear regime, the time evolution 
is examined in section \ref{subsec: time evolution}. 
%
%
%
%
%
%
%
%
%
%
%
%
%
%
%
\subsection{One-loop corrections of power spectra} 
\label{subsec: 1-loop}
%
%
%
%
When we take into account the next-to-leading order terms, 
  the power spectrum can be written as 
\begin{equation}
 P_{(m,g,\times)}(k)= P_{(m,g,\times)}^{(11)}(k)+
  \left[P_{(m,g,\times)}^{(22)}(k)+P_{(m,g,\times)}^{(13)}(k)\right]+\cdots. 
\label{weakly-PS}
\end{equation}
The terms in the bracket are the one-loop 
correction, which can be evaluated using the second and third order 
solutions of perturbations $\dg$ and $\dm$. The details of calculation and the 
resulting expressions for the one-loop spectra $P^{(22)}(k)$ 
and $P^{(13)}(k)$ are comprehensively described in appendix. Here, we discuss 
the technical issue to implement the numerical integration.  

The expressions for the spectra (\ref{Pm22})-(\ref{Pcross13}) 
  have the three-dimensional integral 
  for the linear power spectra $P_l(k)$. 
  The further reduction enables us to obtain the expressions 
  including the two-dimensional integral over $u$ and $x$ for the 
  $P^{(22)}(k)$ part, while the $P^{(13)}(k)$ part has 
  the one-dimensional integral over $x$ 
  (see appendix). The care is required for evaluation of these integrals.  
  For the single power-law spectrum $P_l(k)\propto k^n$ with $n\leq-1$, 
  the infrared divergence appears at each integral. 
  As for the total mass $P^{(22)}_m(k)$ and 
  $P^{(13)}_m(k)$, which have been studied by several authors,  
 the limit $x\to0$ yields
\begin{eqnarray}
  P^{(22)}_m(k),~~P^{(13)}_m(k)~\propto~~k^2 P_l(k) \int_0^{\infty} dq P_l(q), 
\nonumber
\end{eqnarray}
which imply that the infrared divergences occur at $n\leq-1$. 
It is known that this is 
apparent and the divergences cancel out when we evaluate the total 
contribution of the power spectrum $P_m(k)$ (\cite{MSS92},
\cite{JB94}, \cite{SF96b}). However, in cases of 
the power spectra $P_{g}(k)$ and $P_{\times}(k)$, 
the infrared divergences still remain even if we compute the total 
contribution.   

The physical interpretation of the infrared divergence 
in cases with the single power-law spectrum $P_l(k)$ is difficult. 
In reality, however, 
the infrared divergence does not appear if we adopt the phenomenological 
spectrum such as the Cold Dark Matter (CDM) power spectrum 
by Bardeen {\it et al.} (1986), where the effective spectral index  
$n_{eff}\equiv d\log{P_l(k)}/d\log{k}$ becomes positive in the limit 
$k\to0$. Therefore, the particular behavior in the single power-law 
       spectrum  might not affect the real universe. 

On the other hand, depending on the shape of the power spectrum 
      $P_l(k)$, the ultraviolet divergence ($x\to\infty$) 
      appears even in the total mass $P_m(k)$ (\cite{SF96b}). 
      The integrand from the high frequency part can give the 
      large contribution to the total power spectra and  
      the care is also required 
      to proceed the weakly non-linear analysis self-consistently. Thus, 
      the high-frequency cutoff $k_c$ should be introduced when we evaluate 
      the integrals. If we have the power spectrum $P_l(k)$ as 
      defined with the interval $\epsilon \leq k \leq k_c $, 
      the integral over $x$ and $u$ for the $P^{(22)}(k)$ part 
       and over $x$ for the $P^{(13)}(k)$ part should be replaced: 
  \begin{equation}
    \int_0^{\infty}dx \longrightarrow \int_{\epsilon/k}^{k_c/k}dx,
    ~~~~~~~~~~~~~
    \int_{-1}^1 du \longrightarrow 
\int_{\mbox{Max}\left\{-1,(1+x^2-(\epsilon/k)^2)/(2x)\right\}}
^{\mbox{Min}\left\{1,(1+x^2-(k_c/k)^2)/(2x)\right\}} du. 
\nonumber
  \end{equation}
%
The cutoff $\epsilon$ is auxiliary introduced so that 
the numerical integration can converge rapidly. 
As long as the power spectrum $P_l(k)$ has the effective index 
$n_{eff}>-1$ in the limit $k\to0$, the influence of the cutoff $\epsilon$ 
does not affect the weakly non-linear power spectra. 
%
%
%
%
%
%
%
%
\subsection{Non-linear biasing and power spectrum} 
\label{subsec: weak nonlinear} 
%
%
%
%
%
%
The qualitative features 
       of the one-loop power spectra should be mentioned 
       before analyzing the time evolution. 
       Several authors have discussed the influence of 
       non-linear biasing to the two-point correlation function 
       (\cite{Coles93}, \cite{FG93}, \cite{SW98}, \cite{HMV98}, \cite{DL99}). 
       The remarkable features appearing in the spectrum of galaxies 
       are summarized as follows :
\begin{description}
\item[(i)] Constant power on very large scales, 
\item[(ii)] Over a wide range of scales,  
           the non-vanishing parameters $h_1$ and $h_2$ modify the 
           biasing parameter $b_k$ predicted in linear theory.  
\end{description}
%
%
%
%
%

The effect (i) comes from the $P_g^{(22)}(k)$ part 
       of the power spectra. From (\ref{Pg22}), the initial one-loop spectrum 
       $P^{(22)}_g(k)$ becomes 
       \begin{eqnarray}
        \frac{h_1^2}{18}b_0^4~ \frac{k^3}{(2\pi)^2}\int dx~x^2 P_l(kx)~
         \int du ~P_l(k\sqrt{1+x^2-2xu}). 
         \nonumber 
       \end{eqnarray}
       For the typical linear spectrum $P_l(k)$ satisfying the
       limit $P_l(k) \to0$ as $k\to0$, the above integral becomes
       constant and it eventually dominates over both the linear term 
       $P_g^{(11)}$ and $P_g^{(13)}$ part. This leads to the 
       divergence of the biasing parameter $b_k$ on very large scales. 

According to Heavens, Matarrese \& Verde (1998), 
       the constant power in the $P_g^{(22)}$ part is 
       interpreted as the shot-noise term, which is a consequence 
       of the local ansatz for the galaxy distribution 
       (see eq.[\ref{init-g}]). Although the constant power can 
       be measured on extremely large scales, this would become 
       negligible on realistic scales of the galaxy surveys.  
       Therefore, we will analyze the power spectrum $P_g(k)$ 
       on the scales where the effect (i) becomes negligible. 

On the other hand, the clustering of galaxies 
       incorporated by the non-Gaussian initial condition (\ref{init-g})
       leads to the effect (ii), which would have 
       a more practical importance for the data analysis of galaxy surveys. 
       The dominant contribution of this effect 
       comes from the $P^{(13)}(k)$ part. In our case, the initial 
       biasing parameter is estimated using equation (\ref{Pg13}):
  \begin{equation}
    b_k\simeq\sqrt{\frac{P_g^{(11)}+P_g^{(13)}}{P_m^{(11)}}}=
    b_0\left(1+\frac{h_2}{8}b_0^2~\sigma_0^2\right)~~;~~~~
    \sigma_0^2\equiv\int \frac{d^3\q}{(2\pi)^3}~P_l(q). 
    \label{effective-b}
  \end{equation}
%
The expression (\ref{effective-b}) shows that the 
deviation from linear theory prediction depends on the 
parameters  $\sigma_0$, $b_0$ and $h_2$. 
The spectrum $P_l(k)$ with the cutoff parameter implies that 
the parameter $\sigma_0$ would have the cutoff dependence. 
In a realistic situation, the clustering property characterized 
by the assumption (\ref{init-g}) might in general depend on the scale. 
Hence, it is natural to have the cutoff dependence. 
As for the initial correlation coefficient, it remains 
  unchanged due to the assumption (\ref{init-g}).   
The expressions (\ref{Pg13}) and (\ref{Pcross13}) yield 
\begin{eqnarray}
  r_k \simeq \frac{P_{\times}^{(11)}+P_{\times}^{(13)}}
  {\sqrt{(P_g^{(11)}+P_g^{(13)})P_m^{(11)}}}=
  \frac{1+\frac{h_2}{8}~b_0^2\sigma_0^2}{\sqrt{1+\frac{h_2}{4}~b_0^2\sigma_0^2}}
  \simeq r_0+{\cal O}(\sigma_0^4),   
\nonumber
\end{eqnarray}
which is valid as long as $\sigma_0\lesssim1$.  

Notice that the gravitational evolution 
  alters the effective biasing and 
  the correlation coefficient. 
  The non-linear biasing parameters $h_1$ and $h_2$ of 
  the one-loop spectra dynamically affect these parameters. 
  Further, the non-linear growth of 
  density fluctuations might change the evolution.  
%
%
%
%
\subsection{Time evolution of $b_k$ and $r_k$}
\label{subsec: time evolution}
%
%
%
%
%
We now consider the time evolution of the biasing parameter $b_k$ and 
correlation coefficient $r_k$. In the rest of this paper, 
we present the results for the linear spectrum $P_l(k)$ given by 
the standard CDM model. Although we have also examined the two-power-law model 
of Makino, Sasaki \& Suto (1992), we obtained  
the qualitatively similar results to those shown below. 
Thus, our results would be general in the case of the realistic spectrum 
with a break. 

From Bardeen {\it et al.} (1986), the fitting form of 
the linear spectrum $P_l(k)$ becomes
\begin{eqnarray}
  &&P_l(k)=A~k~T^2(k), 
\nonumber \\
&&~~~~~~~~~~~T(k)=\frac{\ln(1+2.34q)}{2.34q}
\left[1+3.89q+(16.1q)^2+(5.46q)^3+(6.71q)^4\right]^{-1/4},
\nonumber \\
&&~~~~~~~~~~~q=\frac{k}{\Gamma~ h~\mbox{ Mpc}^{-1}},  
\label{BBKS-PK}
\end{eqnarray}
where $k$ is in unit of $h~\mbox{Mpc}^{-1}$. We have chosen 
the shape parameter $\Gamma\equiv \Omega_0 h=0.5$, 
corresponding to the parameters $\Omega_0=1$ and  
$H_0=100 h~\mbox{ km s}^{-1} \mbox{Mpc}^{-1}$ with $h=0.5$. 
The linear growing mode $D(t)$ is therefore 
replaced with $a(t)/a(t_i)$. 
The cutoff parameters are specified as $\epsilon=0.0001~ h~\mbox{Mpc}^{-1}$ 
and $k_c=2\pi/8~ h~\mbox{Mpc}^{-1}$, appropriate for examining 
the weakly non-linear evolution on large scales. Then, 
the amplitude $A$ is normalized by $\sigma_8=0.54$ 
at the present time $(z=0)$, determined from the cluster abundance 
(\cite{KS97}). We then perform the numerical integration for 
the one-loop corrected terms. 

%
%
%
%
%
%
The evolved results of the weakly non-linear power spectra 
       (\ref{weakly-PS}) are shown in Figure 1.  The thin-solid, thick-solid, 
       and thick-dashed lines are the power spectra $P_m(k)$, $P_g(k)$ and 
       $P_{\times}(k)$ evaluated at present, respectively. 
       The initial conditions are set at $z=5$. Choosing 
       the non-Gaussian parameters $h_1$ and $h_2$ as zero, 
       we specify the parameters $b_0=3.0,~ r_0=0.8$ so that the 
       initial biasing parameter and initial correlation
       coefficient obtained from (\ref{def-of-b-r2}) become 
       $b_k=3.0$, $r_k=0.8$. 

       In Figure 1, 
       the evolved spectra for galaxies and cross correlation become 
       the larger power than that of the total mass due to the 
       initial large biasing. 
       We have also displayed the linear power spectrum of the
       total mass $P_m^{(11)}(k)$ (see eq.[\ref{linear-PS}]). 
       Compared to the linear order result, 
       the weakly non-linear spectrum shows the enhancement 
       of the power on small scales, while both $P_m^{(11)}(k)$ and 
       $P_m(k)$ have the same power on large scales. This behavior 
       is known as the weakly non-linear effect (\cite{MSS92},
       \cite{JB94}). The validity of the weakly non-linear 
       result has been checked using the non-linear fitting formula 
       by Peacock \& Dodds (1996). The one-loop correction of 
       the total mass 
       rather overestimates the non-linear growth at $z=0$.  
       However, until $z=1$,   
       the weakly non-linear spectrum $P_m(k)$ is well-fitted to the 
       formula by Peacock \& Dodds and 
       the qualitative behaviors did not change at $z=0$. 

In Figure 2, using the definition (\ref{def-of-b-r2}), we evaluate the 
biasing parameter $b_k$ and the correlation coefficient $r_k$ as the 
function of the Fourier mode $k$. 
Within the validity of the perturbation, the snapshots of both 
parameters with the same initial conditions as used in Figure 1 are 
depicted for the different redshift parameters 
$z=5, 3, 2$ and $1$. 
On large scales, the spatial dependence of the biasing parameter and 
the correlation coefficient is very weak 
and they approach the constant values. 
The same behaviors can be obtained in the presence 
of initial non-Gaussianity, consistent with the results by 
several authors (\cite{Man98}, \cite{SW98}, \cite{HMV98}).  
On the other hand, the scale-dependence of the galaxy biasing 
exists on small scales. Although $b_k$ and $r_k$ initially 
lose their power due to the high frequency cutoff $k_c$, 
the spatial variation of the biasing parameter and the 
correlation coefficient gradually changes.  

In the following, we shall separately analyze the time evolution in the linear 
scale and the quasi-linear scale. Here, the 
linear scale means that the evolution of the total mass power spectrum can be 
well-approximated by the linear perturbation (\ref{linear-PS}). The 
quasi-linear scale denotes the scales where the growth of 
non-linear correction for the total mass spectrum can become significant, 
corresponding to $k \gtrsim 0.1~h$ Mpc$^{-1}$ in Figure 1. 
%
%
%
%
%
%
%
%
%
%
%
%
%
\subsubsection{Linear scale}
\label{subsubsec: linear} 
%
%
%
%
%
%
%
%
%
Since the non-locality of the gravitational evolution 
       does not affect the biasing feature in the linear regime, 
       the scale-independence of $b_k$ and $r_k$ simply relies 
       on the assumption of the local initial conditions (see
       eq.[\ref{init-g}]). 
We thus analyze the time evolution by 
       fixing the mode $k=2\pi/100~h$ Mpc$^{-1}$. 
       For our purpose, 
       it is important to clarify 
       the influence of the initial non-Gaussianity 
       and the stochasticity to the evolution of $b_k$ and $r_k$. 
       Hence, by varying the non-Gaussian initial parameters $h_1$ and 
       $h_2$, we consider the two cases : almost deterministic biasing 
       ($b_k=3.0$ $r_k=0.8$ at $z=5$) 
       and stochastic biasing ($b_k=3.0$ $r_k=0.2$ at $z=5$).  
       Because the initial non-Gaussianity slightly affects the 
       initial conditions $b_k$ and $r_k$, 
       we appropriately adjust the model parameters $b_0$ and $r_0$ 
       so that the above conditions are satisfied. 
       The list of the model parameters examined in this 
       subsection is shown in Table 1, together with the evolved 
       results $b_k(z=0)$ and $r_k(z=0)$. They are compared with the 
       linear prediction (\ref{linear-b-r}) under the same initial 
       conditions $b_k$ and $r_k$. The resulting evolution $b_k$
       and $r_k$ is depicted in Figure 3a for 
       the nearly deterministic biasing and Figure 3b for the 
       stochastic biasing case. The results are naively extrapolated 
       to the present time $z=0$.         

First, consider the time evolution of the biasing parameter 
       $b_k$. 
       In the deterministic biasing case (upper panel of Figure 3a),  
       the solid line shows the result with the vanishing initial 
       non-Gaussianity. On the scale $k=2\pi/100~h$ Mpc$^{-1}$,  
       there still exists the scale-dependence of the biasing,  
       as shown in Figure 2.  
       This slightly alters the time evolution of $b_k$, 
       compared to the linear prediction $b^{(l)}$ 
       ({\it thin solid line}). However, this deviation 
       is small and the weakly non-linear evolution 
       is in good accordance with the linear prediction until $z=1$. 
       The same behavior is obtained in the non-Gaussian initial 
       condition $h_2=3.0$ ({\it dotted line}). Although the 
       non-Gaussian parameter $h_2$ modifies the initial biasing 
       parameter $b_0=2.64$ predicted by the linear theory 
       (see eq.[\ref{effective-b}]), 
       the resulting evolution is well approximated by the 
       linear evolution $b^{(l)}$ with the observed 
       initial condition $b_k=3.0$ and $r_k=0.8$. 

       However, the deviation from the linear 
       evolution is found in the presence of the 
       initial non-Gaussianity $h_1\neq0$. The long-dashed and 
       the short-dashed lines are the results with the initial 
       parameter $h_1=3.0$ and $h_1=-3.0$, respectively. 
       In the one-loop power spectrum $P^{(13)}_g(k)$, the terms 
       proportional to the parameter $h_1$ has the time dependence 
       $D^2(t)$ (see eq.[\ref{Pg13}]), which is the same 
       growth rate as appeared in the linear spectrum $P_g^{(11)}$ 
       (see eq.[\ref{linear-PS}]). 
       This means that the clustering property induced by 
       the initial non-Gaussianity tends to remain 
       and the deviation from the linear theory would become 
       increasing. In the $h_1>0$ case, we have the positively skewed 
       distribution of galaxies, while the total mass distribution 
       is nearly Gaussian. Thus, the fluctuation $\dg$ correlated 
       with the total mass becomes more and more concentrated on the 
       total mass, which dynamically leads to the larger biasing 
       parameter $b_k$. Vice versa, the smaller value $b_k$ could be 
       attained in the case of the initial skewness $h_1<0$. 

The results of the stochastic biasing case 
($b_k=3.0$, $r_k=0.2$ at $z=5$) are displayed in Figure 3b. 
       For the biasing parameter $b_k$ with the initial non-Gaussianity 
       $h_1\neq0$, remarkably, the deviation 
       from the linear theory prediction is reduced. 
       From the expressions (\ref{Pg22}) and 
       (\ref{Pg13}), one can show that the one-loop corrected terms 
       proportional to the parameter $h_1$ depends  
       on the multiplicative factor $b_0r_0$. 
       The small value of the initial correlation 
       coefficient $r_k\simeq r_0$ forces the galaxy distribution to 
       correlate with the total mass. This behavior also affects 
       the galaxy clustering and 
       it can weaken the effect of 
       initial non-Gaussianity. Hence, 
       the biasing parameter $b_k$ could be approximated by 
       the linear theory $b^{(l)}$.  

       Next, we consider the correlation coefficient. 
       In the bottom panel of Figure 3, we plot time evolution 
       of $r_k$ with the same initial 
       parameters as examined in the upper panel of Figure 3.  
       As was remarked by Dekel \& Lahav (1999), 
       the non-linear biasing such as the initial condition 
       (\ref{init-g}) can bend the 
       $\dg$-$\dm$ relation and it effectively decreases the correlation 
       coefficient, compared to the linear result $r_k^{(l)}$. 
       In both panels of Figure 3a and Figure 3b, the effect of 
       non-linear biasing, corresponding to the non-vanishing parameters  
       $h_1$ and $h_2$ is very weak. The time evolution of $r_k$ can be 
       well-approximated by the linear prediction with the 
       initial conditions $b_k=3.0, r_k=0.8$ or $b_k=3.0, r_k=0.2$. 

       We have also 
       analyzed the time evolution for more various initial conditions. 
       The examples for the initial anti-biasing case are 
       listed in Table 1. We see that the influence of 
       the initial non-Gaussianity to the parameters $b_k$ and $r_k$ 
       becomes weak and the linear prediction does work well. 
       In Table 1, the cutoff dependence $k_c$ is also examined. 
       The small value of the cutoff wavelength
       $2\pi/k_c=4.35~h$ Mpc$^{-1}$ increases 
       the parameter $\sigma_0$ given by 
       (\ref{effective-b}), which strengthens   
       the efficiency of the non-Gaussianity. 
       In particular, the high frequency cutoff with the 
       initial skewness $h_1=3.0$ has the curious behavior that the 
       present correlation coefficient $r_k$ becomes larger than unity,  
       which cannot be allowed because the Schwarz inequality 
       states $-1<r_k<1$. As for the non-vanishing initial kurtosis $h_2$, 
       we obtain the same present values $b_k$ and $r_k$ as obtained 
       in the cutoff $k_c=2\pi/8~h$ Mpc$^{-1}$ case.  
       However, the effective biasing induced by the non-Gaussianity 
       is significant and we have the small value of the initial 
       parameter $b_0=2.42$. 
%
%
%
%
%
%
%
%
%
%
\subsubsection{Quasi-linear scale}
\label{subsubsec: quasi-linear}
%
%
%
%
%
%
On quasi-linear scales $k\gtrsim0.1~h~$Mpc$^{-1}$, 
the next-to-leading 
order of non-linear gravity gives the important 
contribution to the growth of total matter power spectra 
(see Figure 1). 

We pay attention to the non-linear growth, 
which may be related to the scale-dependence of the biasing shown 
in Figure 2. From the weakly non-linear analysis $P_m(k)$ by several authors, 
we have recognized that the mode-mode coupling of long-wave modes 
contributes to the amplitude of the short-wavelength fluctuations 
(\cite{MSS92}). This leads that the power spectrum evolves with the 
different growth rate, depending on the effective spectral index $n_{eff}$. 
For the typical spectrum like the CDM model, the one-loop correction 
causes the significant enhancement of the high-$k$ part of the power 
spectrum (\cite{JB94}). 

The weakly non-linear effect also appears 
      in the power spectrum of galaxies $P_g(k)$ and the cross 
      correlation $P_{\times}(k)$ (see
      eqs.[\ref{Pg22}]-[\ref{Pcross13}]). 
      On appropriately small scales, we could find 
      the enhancement of perturbation $\dg$, whose growth rate 
      depends on the initial parameters $b_0, r_0$ and the initial 
      non-Gaussianity, in addition to the effective index $n_{eff}$. 
      Hence, the relative differences between the growth rate of 
      $\dg$ and $\dm$ dynamically lead to the scale-dependence of 
      the biasing (see Figure 2). 

To see the effect of non-linear growth explicitly, 
   we write the parameters $b_k$ and $r_k$ as $b_k=b^{(l)}+
   b_k^{loop}$ and $r_k=r^{(l)}+r_k^{loop}$. 
   $b_k^{loop}$ and $r_k^{loop}$ are the one-loop correction for  
   the biasing parameter and the correlation coefficient. They are 
   given by 
\begin{eqnarray}
 b_k^{loop}&=&\frac{b^{(l)}}{2}\left[
\frac{P_g^{(22)}(k)+P_g^{(13)}(k)}{P_g^{(11)}(k)}-
\frac{P_m^{(22)}(k)+P_m^{(13)}(k)}{P_m^{(11)}(k)}\right],
\nonumber\\
 r_k^{loop}&=&r^{(l)}\left[
\frac{P_{\times}^{(22)}(k)+P_{\times}^{(13)}(k)}{P_{\times}^{(11)}(k)}-
\frac{1}{2}\left\{\frac{P_g^{(22)}(k)+P_g^{(13)}(k)}{P_g^{(11)}(k)}+
\frac{P_m^{(22)}(k)+P_m^{(13)}(k)}{P_m^{(11)}(k)}\right\}
\right]. 
\nonumber 
\end{eqnarray}
The scale-dependence induced by gravity can be observed from the 
    time evolution of $b_k^{loop}$ and $r_k^{loop}$. 

      In Figure 4,  
      the snapshots of $b_k^{loop}$ and $r_k^{loop}$ are shown as 
      the function of $k$ for the different redshift $z$. 
      The results at redshift $z=0$ are also displayed to 
      emphasize the qualitative behavior. 
      In any cases, the scale-dependence of the biasing cannot 
      become significant within the validity of the perturbation theory, 
      however, the scale-dependence obviously exists. 
      The variation of the parameters $b_k^{loop}$ and 
      $r_k^{loop}$ becomes larger on smaller scales.   

      The characteristic behavior can be 
      recognized in the two cases $b_0r_0>1$ and $b_0r_0<1$, which can
      be deduced from the one-loop spectrum
      (\ref{Pg22})-(\ref{Pcross13}). 
      For our qualitative understanding, 
      we simply restrict the analysis on the $h_1=h_2=0$ cases. 
      Figure 4a is the case $b_0r_0>1$,  which has the 
      initial condition $b_k=3.0, r_k=0.8$ at $z=5$. 
      Compared to the total mass distribution, the strong clustering 
      of initial galaxies well-correlating with the total mass 
      enhances the non-linear growth of $\dg$. Thus, the one-loop
      correction of the biasing 
      parameter and the correlation coefficient increases in time. 
      On the other hand, the initial distribution 
      $\dg$ with the low correlation does not lead to the enhancement 
      of clustering feature, but rather weakens the growth of fluctuations to 
      have the sufficient correlation with the total mass. 
      Figure 4b corresponds to this situation, in which we have chosen 
      the condition as $b_k=3.0, r_k=0.2$. 
      Because the growth of perturbation $\dm$ becomes 
      larger than that of $\dg$, the biasing parameter $b_k^{loop}$ decreases.  
      Interestingly, the one-loop correction $r_k^{loop}$ initially 
      decreases, although the evolution eventually strengthen 
      the correlation.  
%
%
%
%
%
%
%
%
%
%
%
\section{Summary and Discussion}
\label{sec: conclusion}
%
%
%
%
In this paper, we have focused on the gravitational evolution of the 
stochastic galaxy biasing deduced from the power spectrum. 
Using the perturbation theory, 
we develop the analysis taking into account the 
one-loop correction of the non-linear gravity. Assuming the 
non-local distribution $f(\Dm)$ for the total mass and 
the local initial condition $g(\Dg)$ for the galaxy 
distribution, we examine the weakly non-linear power spectra and 
investigate the time evolution of biasing parameter and the 
correlation coefficient. The important conclusions are summarized as follows: 
\begin{itemize}
 \item  We found that the initial non-Gaussianity affects 
        the galaxy biasing significantly. In the deterministic biasing, 
        the time evolution of the biasing parameter with the 
        non-vanishing initial skewness deviates 
        from the linear theory prediction by 
        Tegmark \& Peebles (1998), even in the linear regime. 
        As time goes on, the deviation cannot be neglected. 
        However, the deviation can be reduced in the presence of the 
        initial stochasticity. On the other hand, the correlation 
        coefficient does not suffer from the initial non-Gaussianity.  
        The time evolution of the correlation coefficient is 
        well-approximated by the linear evolution. 
 \item The local initial condition of the galaxy distribution leads to 
       the almost scale-independent evolution of the biasing on large scales. 
       However, on quasi-linear scales, the one-loop correction of 
       gravity becomes important and it can give the scale-dependence of 
       the biasing. In the typical cases with $b_0r_0>1$ and $b_0r_0<1$, 
       we have the different prediction of the scale-dependence. 
       For the low correlation $b_0r_0<1$, we have the small 
       biasing parameter, compared with the one on large scales. 
\end{itemize}
%
%

      The prediction of the 
      scale-dependent biasing just comes from 
      the weakly non-linear results. We 
      cannot say anything about the strong non-linear regime 
      beyond the one-loop correction. However, the 
      weakly non-linear analysis is suitable in investigating 
      the transition from the linear to the non-linear regime. 
      Therefore, our qualitative 
      prediction will not change as long as the gravity 
      plays a role for the scale-dependent biasing. 
      In fact, the halo biasing property taking into account the merging 
      process has recently shown that the biasing parameter decreases as 
      the smoothing scale of density fluctuation becomes small. 
      (\cite{MW96}, \cite{SL99}, \cite{J98}). 
      Further, there is a suggestion that 
      the halo-halo correlation can become anti-biasing 
      on the relatively large scale $k\sim0.15~h$ Mpc$^{-1}$ 
      (\cite{KK99}). This indicates that the gravitational clustering 
      gives an important contribution. 
      Therefore, the qualitative behavior can be partially understood 
      from our weakly non-linear analysis. 

      As a demonstration, we plot the power spectra $P_m(k)$ and
      $P_g(k)$ at $z=2.2$ and $0.0$ in Figure 5 ({\it left panel}). 
      The parameters are given at $z=2.2$, chosen as $b_k=3.5$,
      $r_k=0.1$ and $h_1=h_2=0.0$, i.e, we have the large scatter 
      in the initial $\dg$-$\dm$ relation. On the scales 
      $k\gtrsim0.6~h$ Mpc$^{-1}$, 
      the perturbation theory breaks down at $z=0.0$ that the
      one-loop correction of the total mass dominates the linear order 
      power spectrum $P_m^{(11)}$. However, within the validity domain, 
      the weakly non-linear spectrum $P_g(k)$ decreases its power in 
      the high frequency part and the anti-biasing property appears.    
      This behavior can be illuminated by comparing with the N-body 
      simulation in the right panel of Figure 5, where the spectrum of 
      dark matter particles $P_m(k)$ and halos $P_h(k)$ are presented.  
      The halo and the dark matter catalogue of the simulation are 
      based on the standard CDM model (\cite{MJS00}). 
      The halos are identified using the Friend-of-Friends 
      algorithm and selected with the mass threshold 
      $M_{th}=4.47\times10^{12}M_{\odot}$. Although the halo biasing 
      at $z=2.2$ is rather larger than the weakly non-linear biasing,  
      the similarities at $z=0.0$ is apparent. For the detailed 
      comparison, further implementation of our analysis is needed, 
      including the merging process 
      of the halos. The volume exclusion of the halos could be 
      essential for the anti-biasing on small scales. However, 
      the similarities seen in Figure 5 implies that 
      the gravitational evolution and the stochasticity of 
      the galaxy biasing play important role for the 
      present biasing features. 
      We will proceed to improve our treatment quantitatively 
      to explain the halo biasing. 

      Throughout this paper, we have neglected the galaxy 
      formation. For a realistic situation including this process, 
      the cosmological simulation by Blanton, Cen \& Tegmark 
      (1999) shows that the hot gas in the overdense region 
      prevents the star formation and the relation between the 
      galaxies and the dark matter becomes a poor correlation. 
      The realistic process will play a role for explaining the 
      origin of stochasticity. Combining it with our analysis 
      might yield the interesting results. 


      Potentially, the effective biasing parameter due to the initial 
      non-Gaussianity can cause a serious problem. 
      The determination of the parameters $h_1$, $h_2$ 
      and the correlation coefficient $r_k$ from the observational data 
      would become difficult 
      if there exists the significant contribution of 
      the non-Gaussianity to the galaxy clustering. 
      The method using the higher order correlation (\cite{F94}, 
      \cite{GF94}) should be used carefully, which is partly 
      mentioned by Taruya \& Soda (1999). 
      This might lead to a more practical issue, 
      that is, the theoretical prediction of the high redshift objects 
      such as the quasars. 
      In this case, the various cosmological effects should be taken 
      into account. The cosmological redshift distortion and the light-cone 
      effect appeared in the statistical quantities 
      are not negligible for the data analysis (\cite{MS96}, \cite{YS99}). 
      The situation will become more complicated in the case of the biasing 
      being stochastic and non-linear. We shall clarify this issue
      in the next task. 
%
%
%
%
%
%

The author would like to thank Y.Suto for careful reading of the 
manuscript and stimulating discussions, and J.Soda for suggesting 
the idea of this work. We are also grateful to H.Magira and Y.P.Jing for 
providing us the data of the simulation catalog. 
%
%
%
%
%
%
%
%
%
%
%
%
%
%
%
%
%
%
%
\newpage
\appendix
\renewcommand{\theequation}{A-\arabic{equation}}
\section*{Appendix: Calculation of the one-loop power spectra}
%
%
In this appendix, we compute the one-loop correction of the power spectra. 

We start to obtain the second and the third order solutions of 
$\dm$ and $\dg$. To get the solutions, it is convenient to 
write down the Fourier transform of the evolution equations 
(\ref{basic-eq1}), (\ref{basic-eq2}) and (\ref{basic-eq-gal}). 
The quantities $\dm$, $\dg$ and $\theta$ are expanded as 
\begin{eqnarray}
 \delta_{m,g}(x,t)=\int \frac{d^3\k}{(2\pi)^3}e^{-i\k\x}~
  \hat{\delta}_{m,g}(\k,t), ~~~~~
 \theta(x,t)=\int \frac{d^3\k}{(2\pi)^3}e^{-i\k\x}~
  \hat{\delta}_{m,g}(\k,t), 
\nonumber
\end{eqnarray}
Then the evolution equations are rewritten with  
\begin{eqnarray}
&& H^{-1}\frac{\partial\hat{\delta}_m(\k,t)}{\partial t}+\hat{\theta}(\k,t)=-
\int\frac{d^3\k'}{(2\pi)^3}{\cal P}(\k',\k-\k')
\hat{\delta}_m(\k-\k',t)\hat{\theta}(\k',t),
\nonumber
\\
&& H^{-1}\frac{\partial\hat{\theta}(\k,t)}{\partial t}+\left(1-\frac{\Omega}{2}+
\frac{\Lambda}{3H^2}\right)\hat{\theta}(\k,t)+\frac{3}{2}\Omega\hat{\delta}_m
(\k,t)
\nonumber \\
&&~~~~~~=-\int\frac{d^3\k'}{(2\pi)^3}\left[{\cal P}(\k',\k-\k')+
{\cal P}(\k-\k',\k')-2{\cal L}(\k',\k-\k')\right]
\hat{\theta}(\k-\k',t)\hat{\theta}(\k',t),
\nonumber
\end{eqnarray}
\begin{eqnarray}
 H^{-1} \frac{\partial\hat{\delta}_g(\k,t)}{\partial t}+\hat{\theta}(\k,t)
=-\int\frac{d^3\k'}{(2\pi)^3}{\cal P}(\k',\k-\k')
\hat{\delta}_g(\k-\k',t)\hat{\theta}(\k',t), 
\nonumber
\end{eqnarray}
where we define 
\begin{eqnarray}
&&{\cal P}(\k_1,\k_2)
=1+\frac{(\k_1\cdot \k_2)}{|\k_1|^2},~~~~~
  {\cal L}(\k_1,\k_2)
=1-\frac{(\k_1\cdot \k_2)^2}{|\k_1|^2|\k_2|^2}. 
\nonumber
\end{eqnarray}
Substituting the linear order solutions (\ref{linear-sol}) into the 
right hand side of the above equations, 
the second order solutions satisfying the initial conditions become 
(\cite{F84}, \cite{TKS99})
\begin{equation}
  \hat{\delta}_m^{(2)}(\k,t)=\int\frac{d^3\k_1d^3\k_2}{(2\pi)^3}
  ~\delta_D(\k-\k_1-\k_2)~\hat{\Delta}_m(\k_1)
\hat{\Delta}_m(\k_2)
\left\{D^2(t)({\cal P}_{1,2}-\frac{3}{2}{\cal L}_{1,2})
+\frac{3}{4}E(t){\cal L}_{1,2}\right\},
\label{2nd-sol-m}
\end{equation}
\begin{eqnarray}
\hat{\delta}^{(2)}_g(\k,t)&=&\hat{\delta}_m^{(2)}(\k,t)-
\hat{\delta}_m^{(2)}(\k,t_i)
\nonumber\\
&+&\frac{1}{6}h_1~\int\frac{d^3\k_1d^3\k_2}{(2\pi)^3}
\delta_D(\k-\k_1-\k_2)
\left\{\hat{\Delta}_g(\k_1)\hat{\Delta}_g(\k_2)
  -(2\pi)^6\langle\hat{\Delta}_g^2\rangle\delta_D(\k_1)\delta_D(\k_2)\right\} 
\nonumber\\
&+&(D(t)-1)\int\frac{d^3\k_1d^3\k_2}{(2\pi)^3}\delta_D(\k-\k_1-\k_2)
{\cal P}_{1,2}
\left\{\hat{\Delta}_m(\k_1)\hat{\Delta}_g(\k_2)
-\hat{\Delta}_m(\k_1)\hat{\Delta}_m(\k_2)\right\}, 
\nonumber\\
\label{2nd-sol-g}
\end{eqnarray}
where the expressions ${\cal P}(\k_1, \k_2)$ and 
${\cal L}(\k_1, \k_2)$ are respectively suppressed as  ${\cal P}_{1,2}$ and 
${\cal L}_{1,2}$. 
The solutions (\ref{2nd-sol-m}) and (\ref{2nd-sol-g}) contain the function 
$E(t)$ which satisfies $E(t_i)=1$. This is the inhomogeneous 
solution of the following equation:
\begin{eqnarray}
&&  \ddot{E}+2H\dot{E}-\frac{3}{2}H^2\Omega E=
3H^2\Omega D^2+\frac{8}{3}\dot{D}^2.    
\nonumber
\end{eqnarray}
In Einstein-de Sitter universe, we have  
\begin{eqnarray}
&& E(t)=\frac{34}{21}D^2(t). 
\nonumber
\end{eqnarray}
It is known that the $\Omega$ and $\Lambda$ dependences of $E/D^2$ 
is extremely weak (\cite{B94}). 
Therefore, we develop the third order perturbations by replacing $E(t)$ with 
$(34/21)D^2(t)$.
Then we obtain the solutions 
\begin{eqnarray}
 \hat{\delta}_m^{(3)}(\k,t)&=&D^3(t)\int\frac{d^3\k_1 d^3\k_2 d\k_3^3}{(2\pi)^6}
  ~\delta_D(\k-\k_1-\k_2-\k_3)
  ~\hat{\Delta}_m(\k_1)\hat{\Delta}_m(\k_2)\hat{\Delta}_m(\k_3)
\nonumber \\
&& \times \left[\frac{7}{18}{\cal P}_{1,2}({\cal P}_{3,12}+{\cal P}_{12,3})
 +\frac{1}{9}({\cal P}_{1,2}-\frac{4}{7}{\cal L}_{1,2})
({\cal P}_{12,3}+{\cal P}_{3,12}-2{\cal L}_{12,3})\right.
\nonumber \\
&& \left.~~~
-\frac{1}{9}{\cal L}_{1,2}({\cal P}_{3,12}+2{\cal P}_{12,3})
\right],
\label{3rd-sol-m}
\end{eqnarray}
\begin{eqnarray}
 \hat{\delta}_g^{(3)}(\k,t)&= &
 \hat{\delta}_m^{(3)}(\k,t)- \hat{\delta}_m^{(3)}(\k,t_i)
\nonumber\\
&+&\frac{1}{24}h_2~\int\frac{d^3\k_1 d^3\k_2 d\k_3^3}{(2\pi)^6}
  ~\delta_D(\k-\k_1-\k_2-\k_3)~
  \hat{\Delta}_g(\k_1)\hat{\Delta}_g(\k_2)\hat{\Delta}_g(\k_3)
\nonumber\\
&+&\int\frac{d^3\k_1 d^3\k_2 d\k_3^3}{(2\pi)^6}
  ~\delta_D(\k-\k_1-\k_2-\k_3)~
\nonumber\\
&\times&\left[{\cal P}_{3,12}\left\{
-\frac{1}{2}(D(t)-1)\left(
{\cal P}_{1,2}+{\cal P}_{2,1}-\frac{4}{7}{\cal L}_{1,2}\right)
\hat{\Delta}_m(\k_1)\hat{\Delta}_m(\k_2)\hat{\Delta}_m(\k_3)\right.\right.
\nonumber\\
&&~~~\left.\left.+\frac{1}{2}(D(t)-1)^2{\cal P}_{1,2}
\left(\hat{\Delta}_m(\k_1)\hat{\Delta}_g(\k_2)
-\hat{\Delta}_m(\k_1)\hat{\Delta}_m(\k_2)\right)\hat{\Delta}_m(\k_3)
\right.\right.
\nonumber\\
&&~~~\left.\left.+\frac{1}{6}h_1~(D(t)-1)
\left(\hat{\Delta}_g(\k_1)\hat{\Delta}_g(\k_2)-
(2\pi)^6\langle\hat{\Delta}_g^2\rangle\delta_D(\k_1)\delta_D(\k_2)
\right)\hat{\Delta}_m(\k_3)
\right\}\right.
\nonumber\\
&&~~~+\left.\frac{1}{2}{\cal P}_{12,3}(D^2(t)-1)
\left({\cal P}_{1,2}+{\cal P}_{2,1}-\frac{4}{7}{\cal L}_{1,2}\right)
\hat{\Delta}_m(\k_1)\hat{\Delta}_m(\k_2)
\left(\hat{\Delta}_g(\k_3)-\hat{\Delta}_m(\k_3)\right)
\right],
\nonumber\\
\end{eqnarray}
where the quantities ${\cal P}_{ij,l}$, 
${\cal P}_{l,ij}$ and ${\cal L}_{ij,l}$ are respectively given by 
\begin{eqnarray}
 {\cal P}_{ij,l}={\cal P}(\k_i+\k_j,\k_l),~~~ 
 {\cal P}_{l,ij}={\cal P}(\k_l,\k_i+\k_j),~~~ 
 {\cal L}_{ij,l}={\cal L}(\k_i+\k_j,\k_l).
\nonumber
\end{eqnarray}
Using the above results, the one-loop correction for the 
power spectra can be obtained by taking the ensemble average 
(\ref{initial-spectra}):
\\
\\
\noindent
\underline{Total mass distribution}

\noindent
The one-loop power spectra $P_m^{(22)}(k)$ and $P_m^{(13)}(k)$ 
are evaluated from the ensemble average 
\begin{eqnarray}
  \langle\hat{\delta}_m^{(2)}(\k_1)\delta_m^{(2)}(\k_2)\rangle
&=&(2\pi)^3\delta_D(\k_1+\k_2)~P_m^{(22)}(k_1),
\label{pm22}
\\
 2~\langle\hat{\delta}_m^{(1)}(\k_1)\delta_m^{(3)}(\k_2)\rangle
&=&(2\pi)^3\delta_D(\k_1+\k_2)~P_m^{(13)}(k_1).
\label{pm13}
\end{eqnarray}
The results are 
\begin{eqnarray}
  P_m^{(22)}(k)&=&\frac{D^4}{2}~\int\frac{d^3\q}{(2\pi)^3}P_l(q)P_l(|\k-\q|)
  ~\ALP^2(\q, \k-\q),
  \label{Pm22}\\
  P_m^{(13)}(k)&=& 2~D^4~P_l(k)\int\frac{d^3\q}{(2\pi)^3}P_l(q)
  ~\A(\k, \q, -\q), 
  \label{Pm13}
\end{eqnarray}
which coincide with the weakly non-linear analysis of 
Makino, Sasaki \& Suto (1992), Jain \& Bertschinger (1994) and 
Scoccimarro \& Frieman (1996b). 
\\
\\
\underline{Galaxy distribution} 

\noindent
The power spectra $P_g^{(22)}(k)$ and $P_g^{(13)}(k)$ are also  
given by the ensemble averages (\ref{pm22}) and (\ref{pm13}) 
replacing the subscript $(_m)$ with $(_g)$. 
The resulting expressions are 
\begin{eqnarray}
  P_g^{(22)}(k)&=&\int\frac{d^3\q}{(2\pi)^3}P_l(q)P_l(|\k-\q|)
\nonumber\\
&\times&
\left[\frac{1}{2}(D^2-1)^2~\ALP^2(\q, \k-\q)
  +(D^2-1)(D-1)(b_0r_0-1)~\ALP(\q, \k-\q)\BET(\q, \k-\q)
\right.
\nonumber\\
&&\left.+(D-1)^2(b_0r_0-1)^2~\GAM(\q, \k-\q)+
  (D-1)^2(b_0^2-2 b_0 r_0+1)~\EPS(\q, \k-\q)
  \right.
\nonumber\\
&&\left.+\frac{1}{3}h_1(D^2-1)(b_0r_0)^2~\ALP(\q, \k-\q)
+\frac{1}{3}h_1 (D-1) b_0^2r_0(b_0-r_0)~\BET(\q, \k-\q)
\right.
\nonumber\\
&&\left.+\frac{1}{18}h_1^2~b_0^4
\right]    
\label{Pg22}
\\
P_g^{(13)}(k)&=&2~P_l(k)\int\frac{d^3\q}{(2\pi)^3}P_l(q)  
\nonumber\\
&\times&
\left[
(D-1+b_0r_0)\left\{(D^3-1)~\A(\k, \q, -\q)-(D-1)~\B(\k, \q, -\q)
\right.\right.
\nonumber\\
&&~~~~~~~~~~~~~~~~~~~~~~~~~~
\left.\left.-(D-1)^2~\C(\k, \q, -\q)+(D^2-1)(b_0r_0-1)\D(\k, \q, -\q)
\right\}\right.
\nonumber\\
&&\left.+(D-1)^2\left\{\left(b_0r_0(D-1)+b_0^2\right)\C(\k, \q, -\q)
+b_0^2(r_0^2-1)\E(\k, \q, -\q)\right\}\right.
\nonumber\\
&&\left.+\left\{\frac{1}{3}h_1~b_0r_0(D-1)+\frac{1}{8}h_2~b_0^2\right\}
\left\{(D-1)b_0r_0+b_0^2\right\}
\right].
\label{Pg13}
\end{eqnarray}
\\
\\
\underline{Cross correlation}

\noindent
The one-loop power spectra for the cross correlation are computed from 
\begin{eqnarray}
    \langle\hat{\delta}_g^{(2)}(\k_1)\delta_m^{(2)}(\k_2)\rangle
&=&(2\pi)^3\delta_D(\k_1+\k_2)~P_{\times}^{(22)}(k_1),
\nonumber 
\\
\langle\hat{\delta}_m^{(1)}(\k_1)\delta_g^{(3)}(\k_2)
+\hat{\delta}_g^{(1)}(\k_1)\delta_m^{(3)}(\k_2)\rangle
&=&(2\pi)^3\delta_D(\k_1+\k_2)~P_{\times}^{(13)}(k_1).
\nonumber
\end{eqnarray}
Then we have 
\begin{eqnarray}
  P_{\times}^{(22)}(k)
  &=&\int\frac{d^3\q}{(2\pi)^3}P_l(q)P_l(|\k-\q|)~\frac{D^2}{2}~\ALP(\q, \k-\q)
  \nonumber \\
  &\times &\left[
    (D^2-1)~\ALP(\q, \k-\q)+(D-1)(b_0r_0-1)~
\BET(\q, \k-\q)+\frac{1}{3}h_1~(b_0r_0)^2\right]
\nonumber\\
  \label{Pcross22}
  \\
  P_{\times}^{(13)}(k)&=&P_l(k)\int\frac{d^3\q}{(2\pi)^3}P_l(q) 
  \nonumber \\
  &\times& \left[
    \left(D(D^3-1)+D^3(b_0r_0+D-1)\right)\A(\k, \q, -\q)
    -D(D-1)\B(\k, \q, -\q)\right.
  \nonumber \\
  &&~~~~~~~~~~~~~~~\left.
    +(b_0r_0-1)D \left\{(D-1)^2\C(\k, \q, -\q)+(D^2-1)\D(\k, \q, -\q)
    \right\}\right.
  \nonumber \\
  &&\left.+D~b_0r_0\left\{\frac{1}{3}h_1b_0r_0(D-1)+\frac{1}{8}h_2b_0^2\right\}
  \right].
  \label{Pcross13}
\end{eqnarray}
\\
\\

The bold-math quantities appeared in the expressions 
(\ref{Pm22})-(\ref{Pcross13}) are defined 
as follows: 
\begin{eqnarray}
  \ALP(\k_1,\k_2)&=&{\cal P}_{1,2}+{\cal P}_{2,1}
  -\frac{4}{7}{\cal L}_{1,2},
\nonumber 
\\
\BET(\k_1,\k_2)&=&{\cal P}_{1,2}+{\cal P}_{2,1},
\nonumber 
\\
\GAM(\k_1,\k_2)&=&{\cal P}_{1,2}{\cal P}_{2,1},
\nonumber 
\\
\EPS(\k_1,\k_2)&=&\frac{1}{2}\left({\cal P}_{1,2}^2+
{\cal P}_{2,1}^2\right),
\nonumber 
\\
\A(\k_1,\k_2,\k_3)&=&\frac{7}{18}\left\{
{\cal P}_{3,12}\left(
{\cal P}_{1,2}+{\cal P}_{2,1} -\frac{4}{7}{\cal L}_{1,2}
\right)
+{\cal P}_{12,3}\left(
{\cal P}_{1,2}+{\cal P}_{2,1} -\frac{8}{7}{\cal L}_{1,2}
\right)
\right\}
\nonumber\\
&&+\frac{1}{9} \left({\cal P}_{12,3}+{\cal P}_{3,12}-2{\cal L}_{12,3}
\right)\left(
{\cal P}_{1,2}+{\cal P}_{2,1} -\frac{8}{7}{\cal L}_{1,2}
\right),
\nonumber
\\
\B(\k_1,\k_2,\k_3)&=&{\cal P}_{3,12}\left({\cal P}_{1,2}+{\cal P}_{2,1}
  -\frac{4}{7}{\cal L}_{1,2}\right),
\nonumber\\
\C(\k_1,\k_2,\k_3)&=&\frac{1}{2}
{\cal P}_{3,12}\left({\cal P}_{1,2}+{\cal P}_{2,1}\right),
\nonumber\\
\D(\k_1,\k_2,\k_3)&=&\frac{1}{2}{\cal P}_{12,3}\left({\cal P}_{1,2}
+{\cal P}_{2,1}-\frac{8}{7}{\cal L}_{1,2}\right),
\nonumber\\
\E(\k_1,\k_2,\k_3)&=&\frac{1}{2}{\cal P}_{1,2}{\cal P}_{3,12}.
\nonumber
\end{eqnarray}
To develop the further evaluation of the one-loop power spectra, 
we write the integrals in spherical coordinates $q$, $\vartheta$ 
and $\phi$, where we choose the polar axis aligned with 
the wave vector $\k$. Then we introduce the new variables 
$x\equiv q/k$ and $u\equiv \cos{\vartheta}$.  
The integral over the azimuthal angle $\phi$ simply 
yields the multiplicative factor $2\pi$ to each one-loop spectrum. 
As for the integration of $u$,  
the $P^{(22)}(k)$ part needs the explicit expression of the linear spectrum 
$P_l(k)$, while we can integrate the $P^{(13)}$ part 
without knowing the function $P_l(k)$. Thus, 
using the explicit form $P_l(k)$ given by (\ref{BBKS-PK}), 
we numerically evaluate the integral over $x$ and 
$u$ for the $P^{(22)}(k)$ part. 
For the $P^{(13)}(k)$ part, the one-dimensional integration 
over $x$ is only required.  
\clearpage
%
%
%

%
%
%
%
%
%
%
%
%
%
%
%
%
%
%
%
%
%
%
%
%
%
%
\newpage
%
%
%
%
%
%
\begin{deluxetable}{lccccccccccc}
\tablecolumns{13}
\tablecaption{The list of model parameters used in Figure 3 
 and the initial conditions $b_k,~r_k$ 
given at $z=5$ for the fixed mode $k=2\pi/100~h$ Mpc$^{-1}$. 
Together with the linear results $b^{(l)}$ and $r^{(l)}$, 
the present parameters $b_k$ and $r_k$ are also given 
by extrapolating the weakly non-linear analysis.}
\tablewidth{0pc}
\tablehead{
   \colhead{cutoff length} & \colhead{} & \multicolumn{4}{c}{model parameters}
 & \colhead{} &  
 \multicolumn{2}{c}{$z=5$} & \colhead{} &  
 \multicolumn{2}{c}{$z=0$} \\
 \cline{3-6} \cline{8-9} \cline{11-12} \\
 \colhead{$2\pi/k_c $~~[$h^{-1}~$Mpc]} & \colhead{} &
 \colhead{$b_0$} & \colhead{$r_0$} & \colhead{$h_1$} & \colhead{$h_2$} & 
 \colhead{} &   
 \colhead{$b_k$} & \colhead{$r_k$} & 
 \colhead{} &   
 \colhead{$b_k~~(b^{(l)})$} & \colhead{$r_k~~(r^{(l)})$} }
\startdata
\phs  8 &&  3.0 & \phs 0.8 &  \phs 0.0 & 0.0 && 
3.00 & 0.80 && 1.44(1.27) & 0.99(0.97) \nl
\phs  8 &&  2.93 & \phs 0.82 &  \phs 3.0 & 0.0 && 
3.00 & 0.80 && 1.69(1.27) & 1.00(0.97) \nl
\phs  8 &&  2.93 & \phs 0.82 &  \phs -3.0 & 0.0 && 
3.00 & 0.80 && 1.13(1.27) & 1.00(0.97) \nl
\phs  8 &&  2.64 & \phs 0.79 &  \phs 0.0 & 3.0 && 
3.00 & 0.80 && 1.40(1.27) & 0.99(0.97) \nl
\cline{1-12}
\phs  8 &&  3.0 & \phs 0.20 &  \phs 0.0 & 0.0 && 
3.00 & 0.20 && 1.03(1.05) & 0.85(0.89) \nl
\phs  8 &&  2.93 & \phs 0.205 &  \phs 3.0 & 0.0 && 
3.00 & 0.20 && 1.08(1.05) & 0.85(0.89) \nl
\phs  8 &&  2.93 & \phs 0.205 &  \phs -3.0 & 0.0 && 
3.00 & 0.20 && 0.97(1.05) & 0.91(0.89) \nl
\phs  8 &&  2.64 & \phs 0.199 &  \phs 0.0 & 3.0 && 
3.00 & 0.20 && 1.02(1.05) & 0.88(0.89) \nl
\cline{1-12}
\phs  8 &&  0.50 & \phs 0.80 &  \phs 3.0 & 0.0 && 
3.00 & 0.80 && 0.85(0.91) & 1.00(1.00) \nl
\phs  8 &&  0.50 & \phs 0.80 &  \phs 0.0 & 3.0 && 
3.00 & 0.80 && 0.85(0.91) & 1.00(1.00) \nl
\phs  4.35 &&  2.91 & \phs 0.824 &  \phs 3.0 & 0.0 && 
3.00 & 0.80 && 1.97(1.27) & 1.05(0.97) \nl
\phs  4.35 &&  2.42 & \phs 0.782 &  \phs 0.0 & 3.0 && 
3.00 & 0.80 && 1.41(1.27) & 0.99(0.97) \nl
\enddata
\end{deluxetable}
%
%
%
%
%
%
%
%
%
%
%
%
%
%
%
%
%
%
%
%
%
%
\newpage
\begin{figure}
\begin{center}
 \leavevmode
\psfig{file=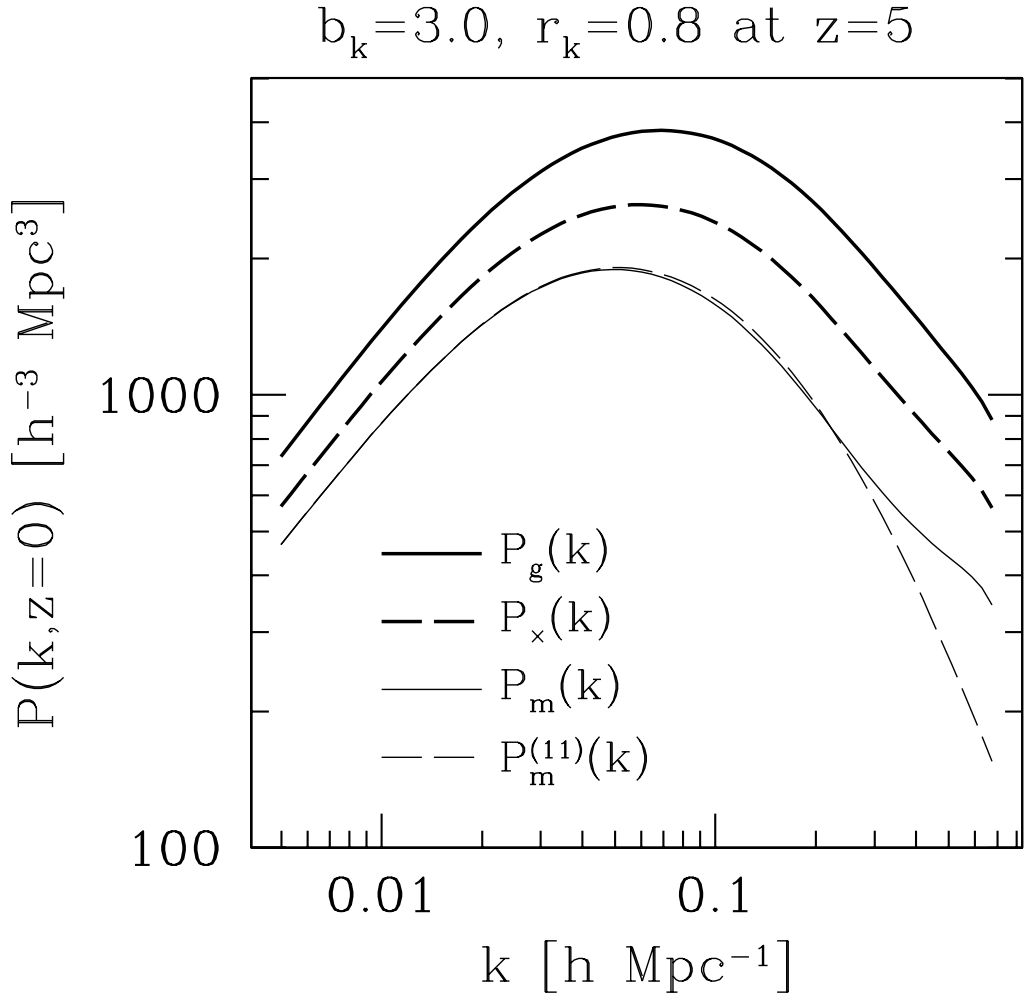,width=16cm} 
\label{spectrum}
\caption{The Evolved results of power spectra at $z=0$.  
            The thin-dashed line 
            shows the linear power spectrum of the total mass
            $P^{(11)}_m(k)$ evolved from $z=5$. 
            The initial spectrum 
            is given by the CDM spectrum (\protect\ref{BBKS-PK}) with the 
            appropriate model parameters (see text). 
            The thin-solid, thick-solid and thick-dashed lines are 
            the weakly non-linear power spectrum of the total 
            mass, galaxies and the cross correlation between them, 
            respectively. The initial conditions at $z=5$ are specified
            as $b_0=3.0$, $r_0=0.8$ and $h_1=h_2=0$. }
\end{center}
\end{figure}
\begin{figure}
\begin{center}
 \leavevmode
\psfig{file=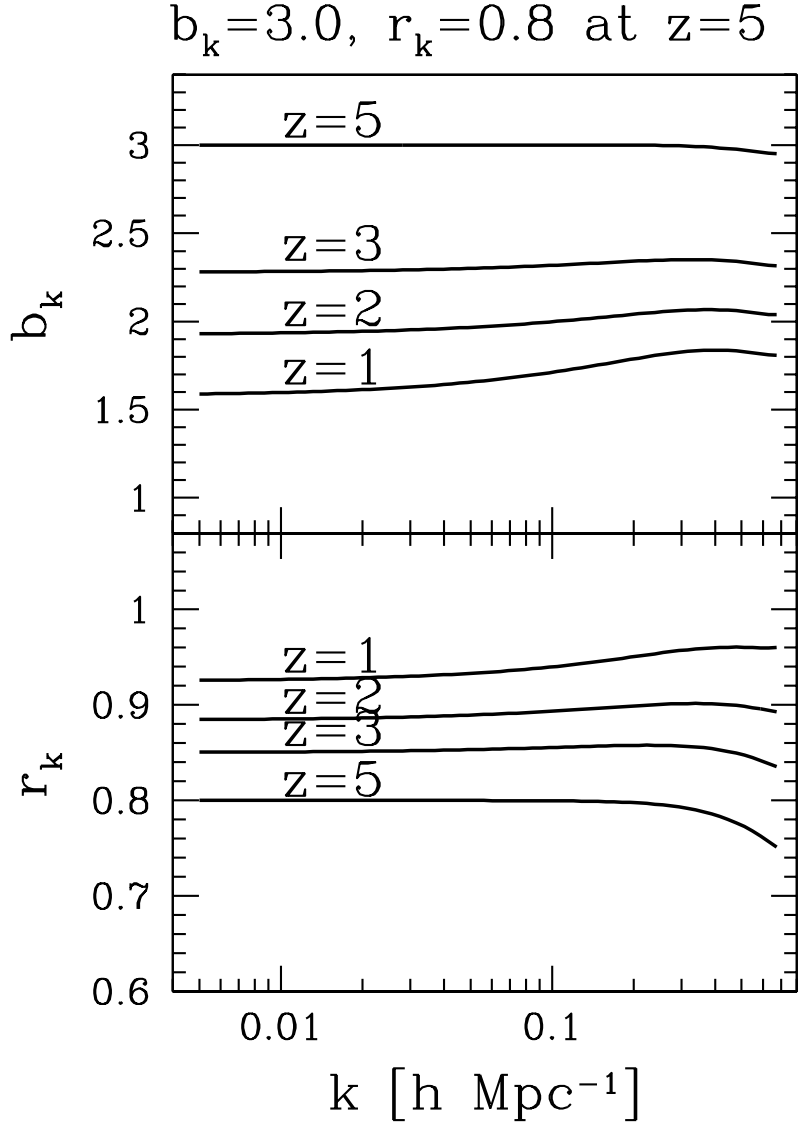,width=16cm} 
\caption{
The snapshots of the evolved biasing parameter({\it upper
            panel}) and the correlation coefficient({\it lower panel}) 
            are depicted as the function of the Fourier mode $k$. 
            The initial condition are the same as in Figure 1. Within 
            the interval $0.001<k<0.1~h~$ Mpc$^{-1}$, the biasing
            parameter and the correlation coefficient becomes almost 
            independent of scales. }
\end{center}
\end{figure}
\clearpage
{
\setcounter{table}{\value{figure}}
\addtocounter{table}{1}
\setcounter{figure}{0}
\renewcommand{\thefigure}{\thetable\alph{figure}}
\begin{figure}
\begin{center}
 \leavevmode
\psfig{file=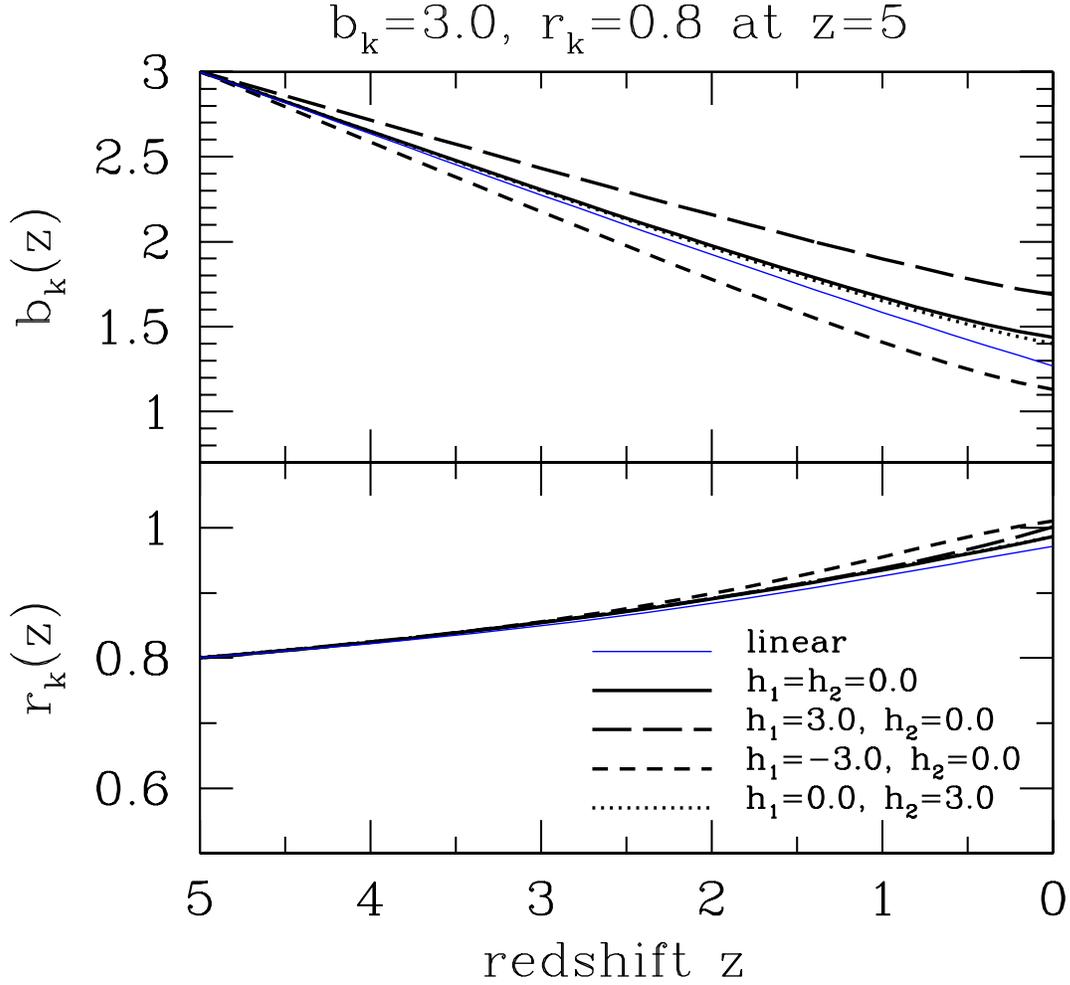,width=15.5cm} 
\end{center}
\caption{Time evolution of the biasing 
            parameter $b_k$ ({\it upper panel}) and the 
            correlation coefficient $r_k$ ({\it lower panel}) 
            as the function of the redshift $z$,  
            fixing the Fourier mode $k=2\pi/100~h~$ Mpc$^{-1}$.   
            By adjusting the model parameters $b_0$ and $r_0$, 
            the initial conditions at $z=5$ are specified as $b_k=3.0$,  
            $r_k=0.8$, i.e, almost {\it deterministic biasing case}.  
            The initial non-Gaussianity are respectively chosen as 
            $h_1=h_2=0.0$({\it solid line}), 
            $h_1=3.0, h_2=0.0$({\it long-dashed line}), 
            $h_1=-3.0, h_2=0.0$({\it short-dashed line}) and  
            $h_1=0.0, h_2=3.0$({\it dotted line}). 
            The results are extrapolated to the present time $z=0$. 
            For comparison, we also plot the linear result given by 
            (\protect\ref{linear-b-r}) 
            with the same initial condition $b_0=3.0$, $r_0=0.8$ 
            ({\it thin-solid line}).}
\end{figure}
\begin{figure}
\begin{center}
\psfig{file=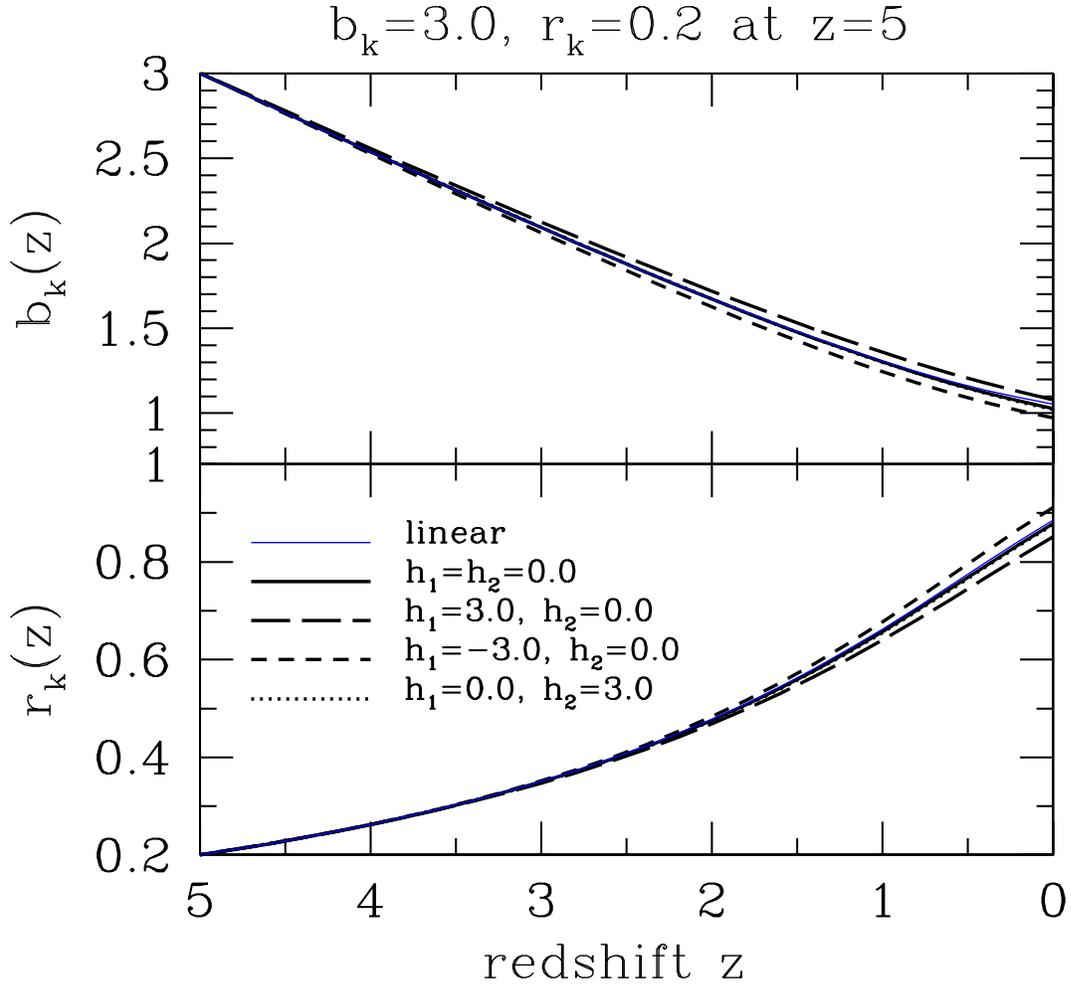,width=15.5cm} 
\label{correl-evolve}
\end{center}
\caption{ The same figure as in Figure 3a, but with the 
different initial conditions $b_k=3.0$, $r_k=0.2$ 
({\it stochastic biasing case}).}
\end{figure}
\setcounter{figure}{\value{table}}
}
\clearpage
{
\setcounter{table}{\value{figure}}
\addtocounter{table}{1}
\setcounter{figure}{0}
\renewcommand{\thefigure}{\thetable\alph{figure}}
\begin{figure}
\begin{center}
 \leavevmode
\psfig{file=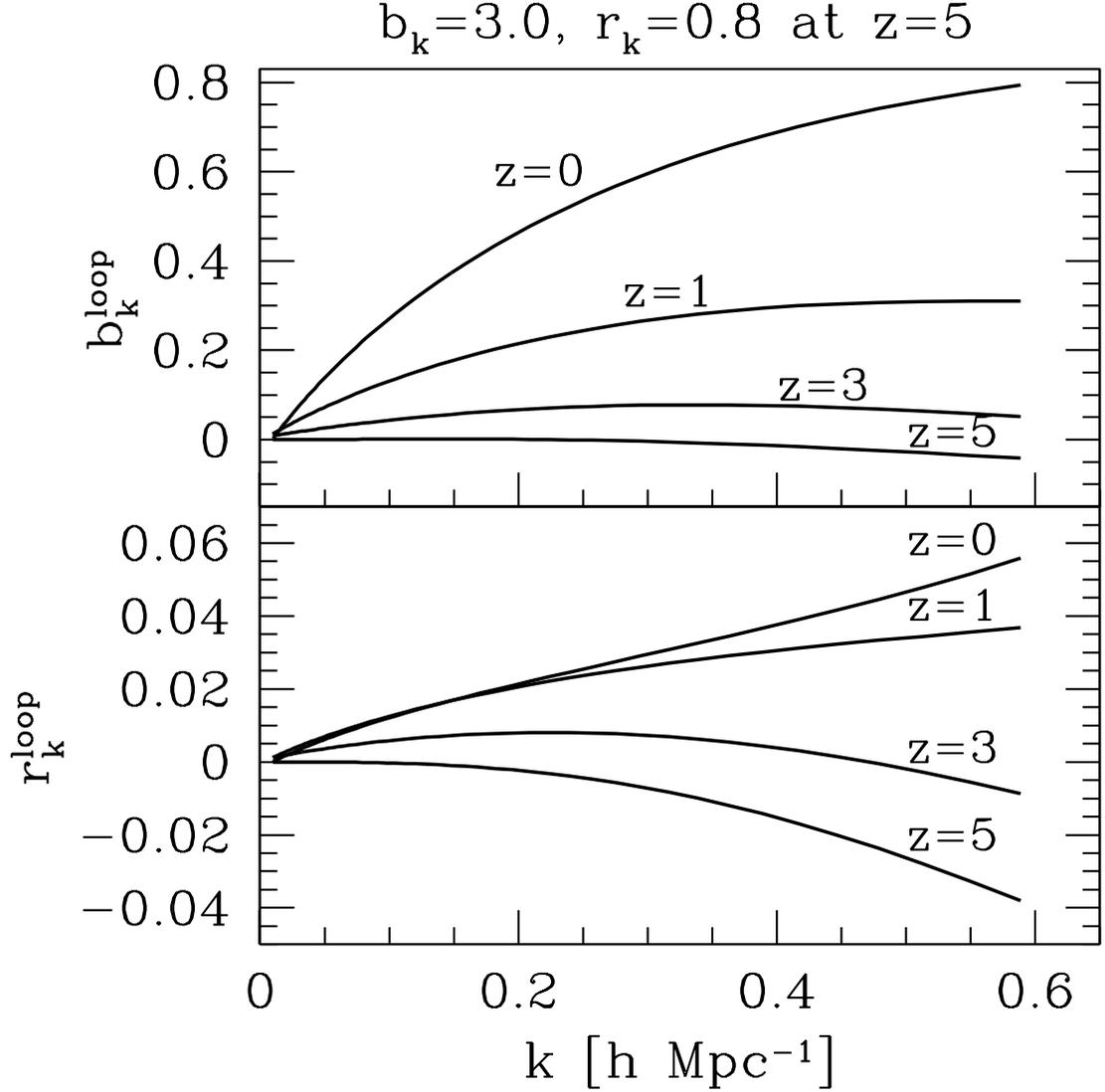,width=16cm} 
\label{comparison}
\end{center}
\caption{The one loop contribution of the biasing 
           parameter $b_k^{loop}$ and the correlation coefficient 
            $r_k^{loop}$ as the function of the 
            Fourier mode $k$. The initial conditions given 
            at $z=5$ are specified as $b_k=3.0$ and $r_k=0.8$.  
            For simplicity, 
            the initial non-Gaussianity are set by $h_1=h_2=0$.  
            Because of the cutoff $k_c$, the initial parameters 
            $b_k^{loop}$ and $r_k^{loop}$ slightly lose their power 
            at high frequency part. However, it is obvious that the spatial 
            dependence of the parameters $b_k^{(loop)}$ and $r_k^{(loop)}$ 
            dynamically appears. On smaller scales, the spatial 
            variation becomes larger. The typical behaviors are
            classified as $b_0r_0>1$ (Figure 4a) and $b_0r_0<1$ 
            (Figure 4b).}
\end{figure}
\clearpage
\begin{figure}
\begin{center}
 \leavevmode
\psfig{file=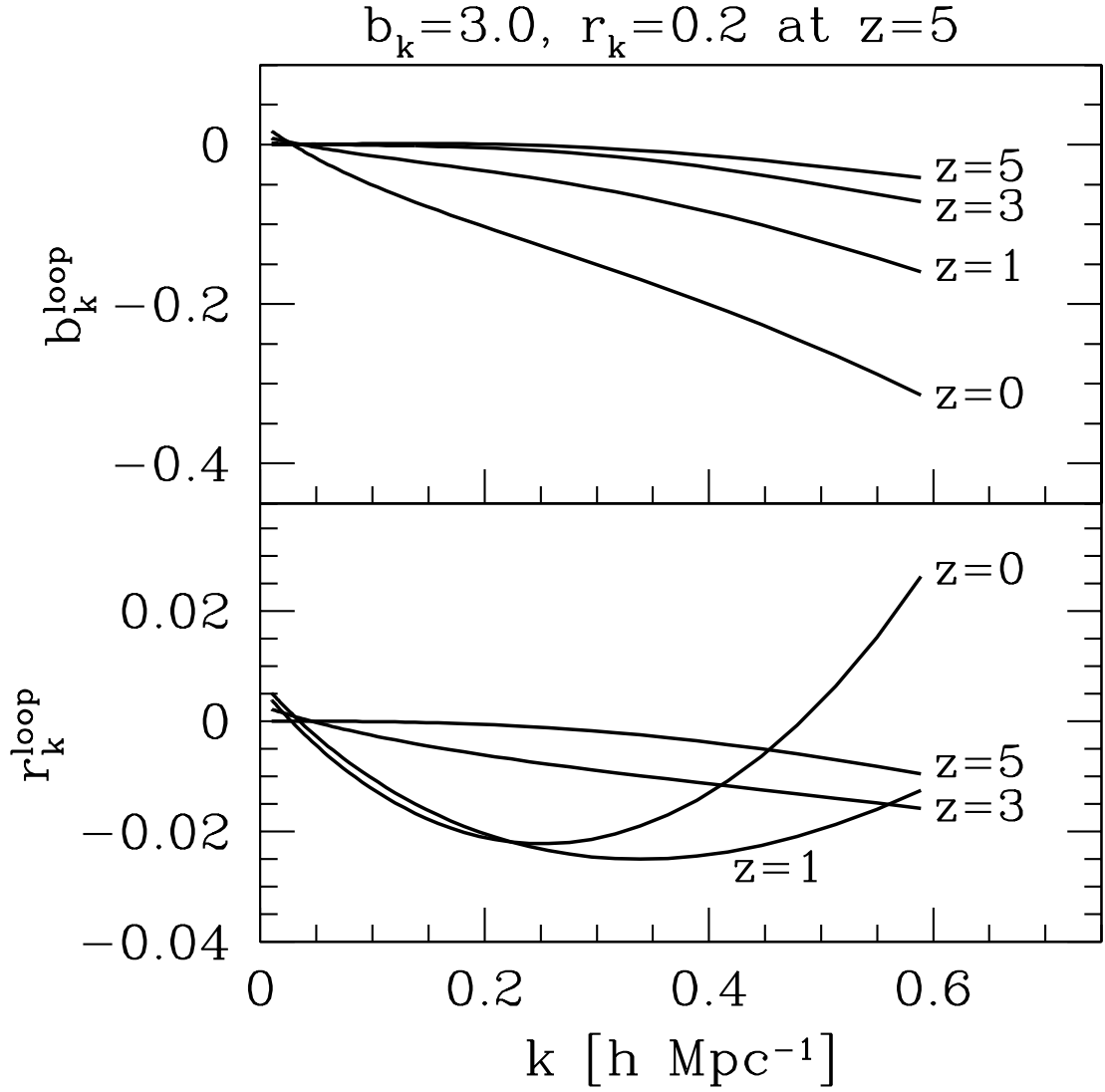,width=16cm} 
\end{center}
\caption{The same figure as depicted in Figure 4a, but with 
            the initial conditions $b_k=3.0$ and $r_k=0.2$, 
            corresponding to the $b_0r_0<1$ case.}
\end{figure}
\setcounter{figure}{\value{table}}
}
\clearpage
\begin{figure}
\begin{center}
 \leavevmode
\psfig{file=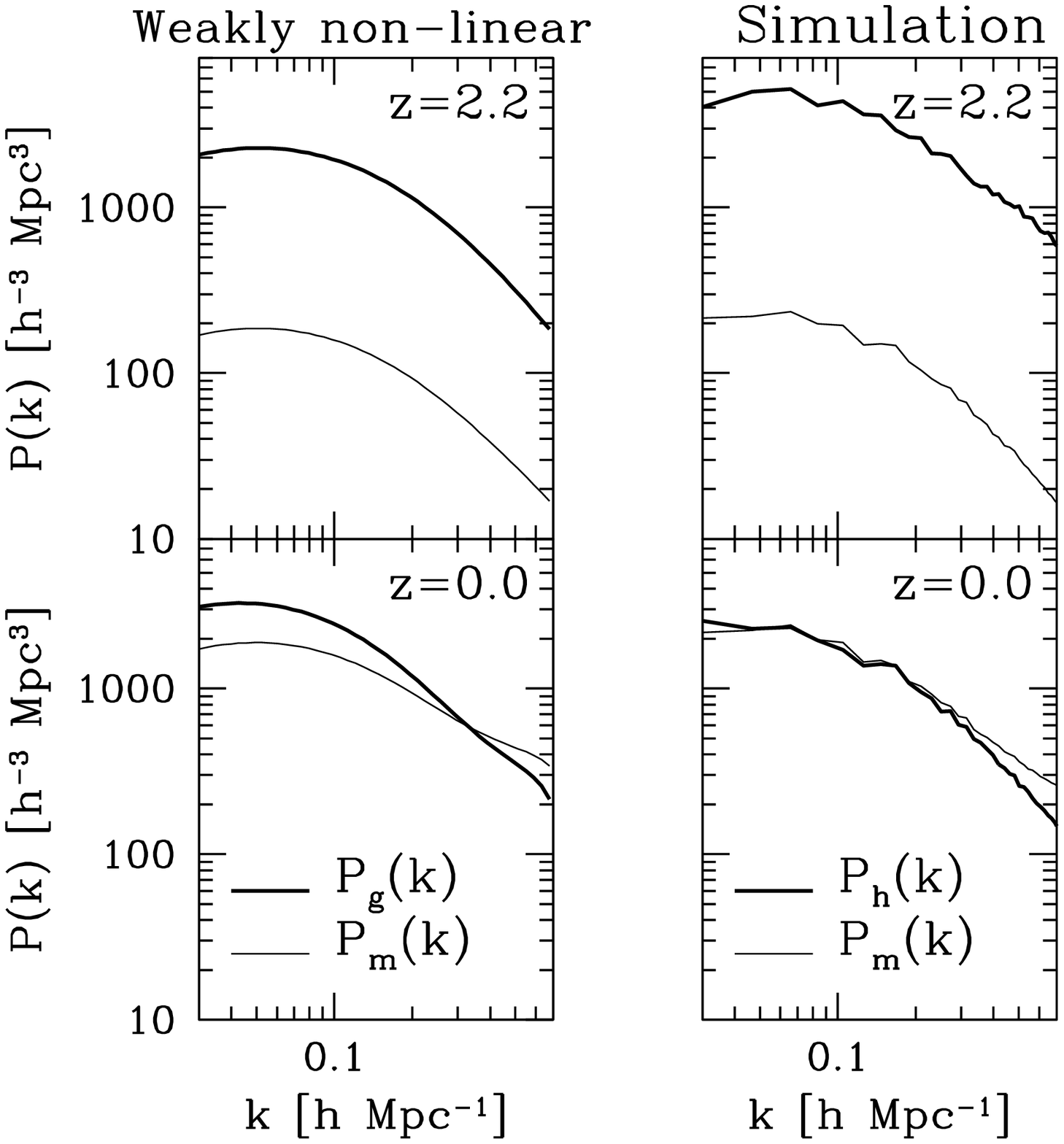,width=17cm} 
\label{demo}
\end{center}
\caption{The anti-biasing property 
in the weakly non-linear analysis ({\it left panel}) and 
 the N-body simulation ({\it right panel}). In left panel, 
 the weakly non-linear power spectra $P_m(k)$ and $P_g(k)$ are 
 depicted under the initial conditions $b_k=3.5$, $r_k=0.1$ and 
 $h_1=h_2=0.0$ given at $z=2.2$. In right panel,  
 using the simulation catalog of Magira, Jing \& Suto (2000), 
 the numerical results of dark matter particles and the halos are 
 presented in the case of Standard CDM model. The halos are identified 
 using the FOF algorithm and selected with the mass threshold 
 $M_{th}= 4.47\times 10^{12}M_{\odot}$. Qualitatively,  
 the weakly non-linear result 
 can recover the anti-biasing feature seen in the simulation 
 if the large scatter in the $\dg$-$\dm$ relation is present initially. }
\end{figure}
\end{document}